\documentclass[twocolumn,aps,showpacs]{revtex4-1}

\usepackage{graphicx}
\usepackage{dcolumn}
\usepackage{amsmath}
\usepackage{color}

\newcommand{\CNPO}{CaNi$_3$P$_4$O$_{14}$}
\newcommand{\CMPO}{CaMn$_3$P$_4$O$_{14}$}

\usepackage{color}
\usepackage[colorlinks,breaklinks,bookmarks=true,citecolor=blue,linkcolor=red,urlcolor=blue]{hyperref}

\begin{document}

\title{Long-range and short-range magnetic correlations, and microscopic origin of net magnetization in the spin-1 trimer chain compound \CNPO}

\author{A. K. Bera}
\email{akbera@barc.gov.in}
\author{S. M. Yusuf}
\email{smyusuf@barc.gov.in}
\author{Amit Kumar}
\affiliation{Solid State Physics Division, Bhabha Atomic Research Centre, Mumbai 400085, India}

\author{M. Majumder}
\altaffiliation{Present address: Max Planck Institute for Chemical Physics of Solids, 01187 Dresden, Germany}
\author{K. Ghoshray}
\affiliation{Condensed Matter Physics Division, Saha Institute of Nuclear Physics, 1/AF Bidhannagar, Kolkata 700064, India}

\author{L. Keller}
\affiliation{Laboratory for Neutron Scattering and Imaging, Paul Scherrer Institut, CH-5232 Villigen PSI, Switzerland}

\date{\today}

\begin{abstract}

Spin-spin correlations and microscopic origin of net magnetization in the spin-1 trimer chain compound \CNPO\ have been investigated by powder neutron diffraction. The present study reveals a three dimensional long-range magnetic ordering below 16 K where the magnetic structure consists of ferromagnetic trimers that are coupled ferromagnetically along the spin-chain direction ($b$ axis). The moment components along the $a$ and $c$ axes arrange antiferromagnetically. Our study establishes that the uncompensated moment components along the $b$ axis ($m_b$) result in a net magnetization per unit cell. The magnetic structure, determined in the present study, is in agreement with the results of recent first principles calculation; however, it is in contrast to a fascinating experimental prediction of ferrimagnetic ordering based on the periodicity of the exchange interactions in \CNPO. Our study also confirms the presence of broad diffuse magnetic scattering, due to 1D short-range spin-spin correlations, over a wide temperature range below $\sim$ 50~K down to a temperature well below the $T_c$. Total neutron scattering analysis by the reverse Monte Carlo (RMC) method reveals that the dominating spin-spin correlation above $T_c$ is ferromagnetic and along the $b$ axis. The nearest neighbor spin-spin correlations along the $a$ and $c$ axes are found to be weakly antiferromagnetic. The nature of the trimer spin structure of the short-range ordered state (above $T_c$) is similar to that of the 3D long-range ordered state (below $T_c$).  The present investigation of microscopic nature of the magnetic ground state also explains the condition required for the 1/3 magnetization plateau to be observed in the trimer spin-chains. In spite of the $S$ = 1 trimer chain system, the present compound \CNPO\ is found to be a good realization of three dimensional magnet below the $T_c$ = 16~K with full ordered moment values of $\sim$ 2~$\mu_B$/Ni$^{2+}$ (1.98 and 1.96 $\mu_B$/Ni$^{2+}$ for two Ni sites, respectively) at 1.5~K.

\end{abstract}


\maketitle

\section{Introduction}
\begin{figure}
\includegraphics[width=85mm]{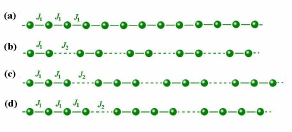}
\caption{\label{Fig:spinchains}(Color online) A schematic representation of (a) linear, (b) dimer, (c) trimer, and (d) tetramer spinchains with defferent periodicity of interactions.}
\end{figure}

The study of quantum spin systems, in particular one dimensional (1D) spin chains, is one of the main interests in condensed matter physics. Magnetic ordering in 1D is generally suppressed even at $T$=0~K by strong quantum fluctuations \cite{HaldanePRL.50.1153,HaldanePLA.93.464}. Moreover, the ground state and low-lying excitations are strongly dependent on the spin value. The ground state of a spin-1/2 Heisenberg antiferromagnetic (AFM) chain belongs to the universality class of Luttinger Liquids, and shows an algebraically decay of correlation function; and magnetic excitations are gapless continuum of spinons \cite{Giamarchi.book}. For the $S$=1 counterpart, the ground state is a nonmagnetic singlet which can be understood in terms of valence bond solid state; and a gap exists between the singlet ground and the first triplet excited states \cite{Bera.PRB.91.144414, Bera.PRB.87.224423}. Long-range magnetic ordering can be stabilized in 1D spin-chain compounds by introducing sufficiently strong interchain interactions and anisotropy \cite{Sakai.PRB.42.4537,Bera.PRB.86.024408,Jain.PRB.87.094411}. Besides uniform spin chains, alternating \cite{Bonner.PRB.27.248}, dimer \cite{Castilla.PRL.75.1823}, trimer \cite{Hase.PRB.74.024430}, tetramer \cite{Hagiwara.JPSJ.72.943} spin-chains [Fig. \ref{Fig:spinchains}] with multiple intra-chain interactions  have recently attracted much attention due to their various unconventional magnetic properties that originate from the periodicity of exchange interactions within the chains.

Trimer spin chains are of interest to us. Trimer spin chains are formed by interconnecting trimers in the one dimension with intratrimer exchange interaction $J_1$, and intertrimer exchange interaction $J_2$, which leads to periodic exchange interactions $J_1$--$J_1$--$J_2$ [Fig. \ref{Fig:spinchains} (c)]. Various magnetic properties are expected depending on the sign and relative strength ($J_2$/$J_1$) of the interactions, spin values, as well as anisotropy. For an AFM trimer chain with an integer spin, the ground state is theoretically predicted to be trimerized; and a gap is present between the ground state and the first excited state for an extended domain of exchange interactions strengths \cite{Schmidt.JPAMT.43.405205}. For the half-integer case, the ground state of an AFM trimer is degenerate \cite{Caspers.PhysicaA.135.519}. However, a rare availability of the real compounds with trimer spin chains limits the experimental investigations, and verification of the theoretical proposals. Nevertheless, interesting macroscopic quantum phenomenon of a 1/3 magnetization plateau was reported experimentally for a few Cu-based spin-1/2 AFM trimer chain compounds, Cu$_3$(P$_2$O$_6$OH)$_2$ \cite{Hase.PRB.73.104419,Hase.PRB.76.064431}, and $A_3$Cu$_3$(PO$_4$)$_4$ ($A$=Ca, Sr, and Pb) \cite{Belik.JSSC.178.709}.

Recent discovery of the trimer spin chain compounds with the general formula $AM_3$P$_4$O$_{14}$ (where $A$=Ca, Sr, Ba, and Pb and $M$=Co, Mn, and Ni) \cite{Yang.IC.47.2562,Lii.IC.32.4373,Elmarzouki.JSP.118.202,Bolte.ACe57.i30,Brahim.SSS.3.669} allows the role of spin value, anisotropy as well as the sign and relative strength ($J_2$/$J_1$) of the interactions to be investigated. The compounds with Mn ($S$=5/2) i.e., $A$Mn$_3$P$_4$O$_{14}$ ($A$=Sr, and Ba) exhibit both quantum-mechanical 1/3 magnetization plateau, and classical magnetic long-range order \cite{Hase.PRB.80.054402,Yang.IC.47.2562}. The magnetic structure is reported to be a coplanar spiral antiferromagnet below $T_N$ (= 2.2 K) \cite{Hase.PRB.84.184435}. Whereas, no magnetization plateau is found for the isostructural Co ($S$ = 3/2) and Ni ($S$ = 1) based compounds having low spin values \cite{Hase.JPSJ.81.064702,Hase.PRB.74.024430}.  Determination of microscopic spin structure is, therefore, demanded on isostructural compounds to understand the reason behind the absence or presence of the magnetization plateau in trimer spin-chain compounds. For SrCo$_3$P$_4$O$_{14}$, a canted AFM ground state is reported below $T_N$ $\sim$ 6.5~K \cite{Hase.JPSJ.81.064702}. However, the exact nature of the magnetic ground state (microscopic magnetic structure) of \CNPO\ is unknown; and it is the subject of the present study.

For the compound \CNPO, there is a fascinating prediction that this compound shows a unique long-range ferrimagnetic ordering due to the periodicity of the exchange interactions \cite{Hase.PRB.74.024430, Majumder.PRB.91.104422}. The prediction is based on Monte Carlo (MC) analysis of bulk magnetization studies. Trimer units in \CNPO\ are described by an AFM intratrimer exchange interaction $J_1$, and a ferromagnetic (FM) inter-trimer exchange interaction $J_2$ along the  spin chain \cite{Hase.PRB.74.024430}. The spontaneous magnetization is explained qualitatively on the basis of a ferrimagnetic long-range order \cite{Hase.PRB.74.024430}. In contradiction to the above proposal, recent first-principles calculation on the same compound estimates that both the $J_1$ and $J_2$ interactions are FM \cite{Majumder.PRB.91.104422}. Recent experimental study by nuclear magnetic resonance (NMR) reveals a two-sub-lattice AFM order \cite{Majumder.PRB.91.104422}. However, neither the first-principles calculation nor the NMR study throws light on the magnetic ground state to explain the origin of the net magnetization. An experimental study to reveal microscopic spin structure is, therefore, essential; and is the subject of the present work. In addition, an unusual weak exponent in the power law behavior of $\frac{1}{T_1}$ below 50~K is also reported from the NMR study \cite{Majumder.PRB.91.104422}, suggesting the presence of short-range spin-spin correlations in \CNPO\ which awaits for an experimental verification.

In the present work, we have employed neutron diffraction to investigate the magnetic ground state and
the microscopic origin of the net magnetization in \CNPO. Our investigation reveals that the magnetic structure consists of FM trimers which are arranged ferromagnetically along the chain ($b$ axis). The present experimental results are consistent with the  first-principles calculation which estimates both FM $J_1$ and $J_2$. The present study also establishes that the net magnetization originates from the uncompensated moment components along the chain direction ($b$ axis), in contrast to the reported ferrimagnetic ordering. Present neutron diffraction study also confirms the presence of short-range spin-spin correlations below $\sim$ 50~K over a wide temperature range down to a temperature well below the 3D long-range magnetic ordering temperature, $T_c$=16~K. Short-range spin-spin correlations have been quantified by total scattering analysis of the neutron diffraction data using the reverse Monte Carlo (RMC) method. Dominant spin-spin correlations above $T_c$ are found to be FM, and along the chain axis ($b$ axis).  \\

\section{Experimental details}

Polycrystalline samples of \CNPO\ were synthesized using a solid state reaction method in air.  Stoichiometric mixture of NiO (99.99 $\%$), CaCO$_3$  (99.99~$\%$) and (NH$_4$)$_2$HPO$_4$ (99.99~$\%$) was heated at 1000~$^\circ$C for 150~hrs. with several intermediate grindings.

Powder neutron diffraction measurements were performed down to 1.5~K. Room temperature diffraction pattern over a wide $Q$-range of 0.5$-$9.5~\AA$^{-1}$ was recorded using the neutron powder diffractometer-II ($\lambda$ = 1.2443~\AA) at Dhruva reactor, Trombay, India. Low temperature (over 1.5--100~K) diffraction patterns were recorded using the DMC diffractometer ($\lambda$ = 2.4585~\AA) at the Paul Scherrer Institute (PSI), Switzerland. The measured diffraction patterns were analyzed using the Rietveld refinement technique (by employing the FULLPROF computer program \cite{Fullprof}). Short-range spin-spin correlations were investigated by the RMC simulation based program RMCPOW \cite{RMCPOW}.\\

\section{Results and discussions}

\subsection{Crystal structure}
\begin{figure}
\includegraphics[clip=true, width=85mm]{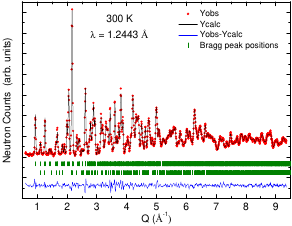}
\caption{\label{Fig:refined_pattern-300K}(Color online) Experimentally observed (circles) and calculated (solid line through the data points) neutron diffraction patterns for \CNPO\ at 300 K. The difference between observed and calculated patterns is shown by the solid line at the bottom. The vertical bars indicate the positions of allowed nuclear Bragg peaks for the main phase \CNPO\ (top row) and the minor secondary phase Ni$_3$P$_2$O$_8$ (bottom row), respectively.}
 \end{figure}

\begin{figure}
\includegraphics[width=87mm]{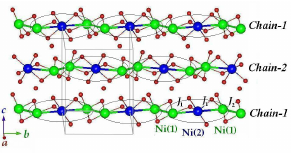}
\caption{\label{Fig:Crystal_Structure_CNPO}(Color online) The crystal structure of \CNPO. The trimer units  are formed by Ni(1)-Ni(2)-Ni(1); as marked by ellipsoids.}
 \end{figure}

The crystal structure of \CNPO\ is investigated by neutron diffraction at room temperature. The Rietveld refined neutron diffraction pattern (measured at Trombay) is shown in Fig.~\ref{Fig:refined_pattern-300K}. The Rietveld analysis confirms that \CNPO\ crystallizes in the monoclinic symmetry (space group: $P2_1/c$) with lattice parameters $a$~=~7.3091(8)~\AA, $b$~=~7.5574(9)~\AA, $c$~=~9.3545(11)~\AA, and $\beta$~=~111.989(7)$^\circ$ at 300~K; which are consistent with the previous reports \cite{Hase.PRB.74.024430}. The refined values of fractional atomic coordinates and isotropic thermal parameters are given in Table \ref{T:structure}. All the atomic sites are considered to be fully occupied, and kept fixed during the refinement.

\begin{table}
\caption{\label{T:structure}The Rietveld refined fractional atomic coordinates, and isotropic thermal parameters ($B_{iso}$) for \CNPO\ at room temperature. $Occ.$ stands for site occupancy.}
\begin{ruledtabular}
\begin{tabular}{ccccccc}

Atom&Site&$x/a$&$y/b$&$z/c$&$B_{iso}$&$Occ.$\\
\hline
Ca & 2$b$& 0.5& 0&0&0.64(4)&1.0\\
Ni(1) & 4$e$& 0.1840(6)& 0.6250(6)&0.0237(4)&0.69(6)&1.0\\
Ni(2) & 2$a$& 0& 0&0&0.57(3)&1.0\\
P1 & 4$e$& 0.8794(11)& 0.7021(12)&0.1910(9)&0.62(2)&1.0\\
P2 & 4$e$& 0.6048(11)&0.4333(10)&0.2093(9)&0.51(4)&1.0\\
O1 & 4$e$& 0.3921(10)&0.4778(10) &0.1963(8) &1.94(6)&1.0\\
O2 & 4$e$& 0.0106(9)&0.2098(8)&0.1346(8)&0.68(4)&1.0\\
O3 & 4$e$& 0.9728(10)&0.5781(10)&0.1171(8)&0.79(5)&1.0\\
O4 & 4$e$& 0.6731(9)&0.6247(10)&0.1734(8)&0.85(3)&1.0\\
O5 & 4$e$& 0.6196(10)&0.2999(10)&0.0868(8)&0.79(4)&1.0\\
O6 & 4$e$& 0.2718(9)&0.8832(10)&0.1323(8)&1.08(5)&1.0\\
O7 & 4$e$& 0.8285(11)&0.8801(10)&0.1033(8)&1.04(4)&1.0\\
\end{tabular}
\end{ruledtabular}
\end{table}

\begin{table*}
\caption{\label{T:bonds} Possible pathways for intratrimer and intertrimer exchange interactions $J_1$ and $J_2$. The Ni--Ni direct distances, bond lengths and bond-angles for the exchange interactions $J_1$ and $J_2$ in \CNPO\ at 300~K.}
\begin{ruledtabular}
\begin{tabular}{ccccc}
Exchange & Ni..Ni direct & Bond lengths & Bond angles  \\
interaction & distance (\AA) & (\AA) & ($^\circ$) \\
\hline\\
$J_1$ & Ni(1)--Ni(2) = 3.109(5) & Ni(1)--O2 = 2.049(7); Ni(2)--O2 = 2.008(7) & Ni(1)--O2--Ni(2) = 100.1(3)\\
& & Ni(1)--O6 = 2.182(8); Ni(2)--O6 = 2.101(6) & Ni(1)--O6--Ni(2) = 93.0(3)\\ \\
$J_2$ & Ni(1)--Ni(1) = 3.179(6) & Ni(1)--O3 = 2.067(8), 2.069(10) & Ni(1)--O3--Ni(1) = 100.5(4)\\
\end{tabular}
\end{ruledtabular}
\end{table*}

The crystal structure of \CNPO\ consists of an edge shared NiO$_6$ octahedra; forming zigzag chains along the crystallographic $b$ axis [Fig. \ref{Fig:Crystal_Structure_CNPO}]. The magnetic Ni$^{2+}$ (3$d^8$, $S$ = 1) ions are located at two crystallographic independent sites, 4$e$ [Ni(1)] and 2$a$ [Ni(2)]. Two types of Ni$-$Ni bonds, viz. Ni(1)$-$Ni(2) and Ni(1)$-$Ni(1) are formed per chain (Fig. \ref{Fig:Crystal_Structure_CNPO}). These two bonds, having different nature [Table \ref{T:bonds}], give rise to the exchange interactions $J_1$ and $J_2$ with different strengths. This results into a spin-1 trimer ($J_1$-$J_1$-$J_2$) chain structure along the $b$ axis. There are two different types of such chains depending on the local orientations of trimers within the chains [marked as chain-1 and chain-2 in Fig. \ref{Fig:Crystal_Structure_CNPO}]. For chain-1, the trimers are slightly deviated (from the $b$ axis) along the negative $a$ axis, whereas, in chain-2, they are deviated along the positive $a$ axis.  The chains are interconnected by P$_2$O$_7$ groups (formed by two corner shared PO$_4$ tetrahedra) in a three-dimensional structure.

The Rietveld analysis also shows the presence of a secondary phase of Ni$_3$P$_2$O$_8$ (weight $\%$ $\approx$ 10 $\%$). The refinement was performed with a two phase model that gives the best agreement between observed and calculated patterns ($R_p$: 2.79~$\%$, $R_{wp}$: 3.67~$\%$, $R_{exp}$: 1.60~$\%$, $\chi^2$: 5.28~$\%$). The crystal structure of the secondary phase Ni$_3$P$_2$O$_8$ is also monoclinic with space group $P2_1/c$ \cite{Escobal.JSSC.178.2626}. The lattice parameters of  Ni$_3$P$_2$O$_8$ are found to be $a$~=~5.7665(5)~\AA, $b$~=~4.6884(4)~\AA, $c$~=~10.1339(8)~\AA, and  $\beta$~=~90.02(9)$^\circ$, which are in good agreement with the values reported in the literature \cite{Escobal.JSSC.178.2626}.

\subsection{Long-range magnetic correlations}
\subsubsection{Magnetic ground state}

\begin{figure}
\includegraphics[clip=true, width=85mm]{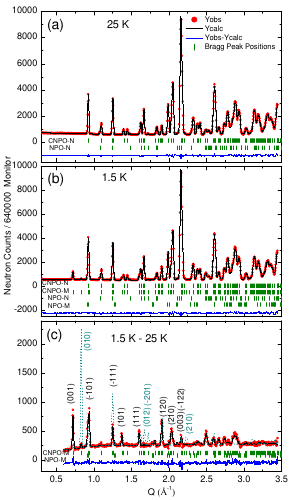}
\caption{\label{Fig:refined_pattern-1p5-25K}(Color online) Experimentally observed (circles) and calculated (solid lines through the data points) neutron diffraction patterns for \CNPO\ at (a) 25~K  (paramagnetic state) and (b) 1.5~K (magnetically ordered state), respectively. (c) Magnetic pattern at 1.5~K (after subtraction of nuclear background at 25~K). The solid lines at the bottom of the each panel represent the difference between observed and calculated patterns. The vertical bars indicate the positions of allowed nuclear (N) and magnetic (M) Bragg peaks for both the main \CNPO\ (CNPO) and impurity Ni$_3$P$_2$O$_8$ (NPO) phases (see text). The dashed line in (c) is the calculated pattern for the AFM trimer model (see text for details).}
\end{figure}

The nature of the magnetic ground state is investigated by neutron diffraction at several temperatures below and above the magnetic ordering temperature, $T_c$ = 16~K. For this purpose, we employed the cold neutron powder diffractometer, DMC at PSI, Switzerland, which provides good intensity as well as good resolution over low $Q$ region. The measured diffraction patterns at 25~K (paramagnetic state) and at 1.5~K (magnetically ordered state) are shown in Fig.~\ref{Fig:refined_pattern-1p5-25K}(a) and Fig.~\ref{Fig:refined_pattern-1p5-25K}(b), respectively. Owing to the presence of a minor Ni$_3$P$_2$O$_8$ secondary phase, the neutron diffraction pattern at 25~K was refined with a two phase model (i.e., the nuclear phases of \CNPO\ and Ni$_3$P$_2$O$_8$). The monoclinic crystal structures with space group $P2_1/c$ for both the nuclear phases (as found at room temperature) reproduce the observed diffraction pattern very well. The diffraction pattern at the base temperature of 1.5~K (magnetically ordered state) shows the presence of a set of additional magnetic Bragg peaks (forbidden in the space group $P2_1/c$) as well as the increase of the intensity of few nuclear Bragg peaks [Fig. \ref{Fig:refined_pattern-1p5-25K}(b)]. All the magnetic peaks could be indexed with a propagation vector ${\mathbf k}$ = (0 0 0) with respect to the monoclinic unit cell of the nuclear phase of \CNPO. In addition, few other weak peaks appear from the AFM [propagation vector = (1/2 1/2 1/2)] phase of Ni$_3$P$_2$O$_8$ ($T_N \sim$ 17.1 K) \cite{Escobal.JSSC.178.2626}.

To determine the symmetry allowed magnetic structure of \CNPO, we performed a representation analysis \cite{Bertaut.JAP.33.1138,Bertaut.ACa.24.217,Bertaut.JPC.32.C1-462,Bertaut.JMMM.24.267,Izyumov-book,Bradley_Cracknell,Cracknell,Wills.PRB.63.064430} using the version 2K of the program SARAH REPRESENTATIONAL ANALYSIS \cite{Wills.physicsB.276.680}.  The inputs for this analysis are the crystal structure above the $T_c$, and the propagation vector of the magnetic ordering. The magnetic representation of a crystallographic site can then be decomposed in terms of the irreducible representations (IRs) of $G_{\mathbf k}$ as

\begin{equation}
 \Gamma_{Mag}=\sum_{\nu} n_{\nu}\Gamma_{\nu}^{\mu}
 \label{magnetic_representation}
\end{equation}

\noindent where n$_\nu$ is the number of times that the IR $\Gamma_\nu$ of order $\mu$ appears in the magnetic representation $\Gamma_{Mag}$ for the chosen crystallographic site. The number of possible "symmetry-allowed'' magnetic structures is simply the number of nonzero IRs in the magnetic representation.

\begin{table}
\caption{\label{tab:IR}Irreducible representations of the group  $G_{\mathbf k}$ of the propagation vector ${\mathbf k}$ = (0 0 0) for \CNPO. }
\begin{ruledtabular}
\begin{tabular}{ccccc}
IRs & \multicolumn{4}{c} {Symmetry element of $G_k$} \\
& $\{E |~0~0~0\}$ & $\{C_{2y} |~0~0.5~0.5\}$ & $\{I |~0~0~0\}$ & $\{\sigma_{y} |~0~0.5~0.5\}$ \\
\hline
$\Gamma_1^1$ & 1 & 1 & 1 & 1 \\ \\
$\Gamma_2^1$ & 1 & 1 & -1 & -1 \\ \\
$\Gamma_3^1$ & 1 & -1 &  1 & -1 \\ \\
$\Gamma_4^1$ & 1 & -1 & -1 &  1
\end{tabular}
\end{ruledtabular}
\end{table}

For \CNPO, the crystal structure is monoclinic with space group $P2_1/c$. All the four symmetry operations of this space group leave the propagation vector {\bf k} invariant.  For the propagation vector ${\mathbf k}$ = (0 0 0), the irreducible representations of the little group $G_{\mathbf k}$ are given in Table \ref{tab:IR}. There are four possible IRs which are one dimensional. The magnetic reducible representation $\Gamma_{mag}$ for the Ni(1) (4$e$ site) and Ni(2) (2$a$ site) sites can be decomposed as a direct sum of IRs, i.e.,

\begin{equation}
\label{eq:gamma_Ni1}
\Gamma_{Mag}^{\text{Ni(1)}}=3\Gamma_{1}^{1}+3\Gamma_{2}^{1}+3\Gamma_{3}^{1}+3\Gamma_{4}^{1}
\end{equation}

\noindent and

\begin{equation}
\label{gamma_Ni2}
\Gamma_{Mag}^{\text{Ni(2)}}=3\Gamma_{1}^{1}+0\Gamma_{2}^{1}+3\Gamma_{3}^{1}+0\Gamma_{4}^{1}
\end{equation}

\noindent respectively. The basis vectors (the Fourier components of the magnetization) for the two magnetic sites Ni(1) [4$e$ ($x$, $y$, $z$); (0.1828, 0.6235, 0.0246)] and Ni(2) [2$a$ (0, 0, 0)]  are given in Table \ref{tab:basis_vectors} for all nonzero IRs. The basis vectors are calculated using the projection operator technique implemented in SARAH \cite{Wills.physicsB.276.680, Kovalev}.

\begin{table}
\caption{\label{tab:basis_vectors}Basis vectors of the magnetic sites Ni(1) and Ni(2) with ${\mathbf k}$ = (0 0 0) for \CNPO. Only the real components of the basis vectors are presented. The atoms of the nonprimitive basis are defined according to Ni(1)-1:(0.1828, 0.6235, 0.0246); Ni(1)-2:~(0.8172, 0.1235, 0.4754); Ni(1)-3:~(0.8172, 0.3765, 0.9754); Ni(1)-4:~(0.1828, 0.8765, 0.5246) and Ni(2)-1:~(0, 0, 0); Ni(2)-2:~(0, 0.5, 0.5).}
\begin{ruledtabular}
\begin{tabular}{cccccccccc}
IRs &&&\multicolumn{7}{c}{Basis Vectors} \\
\cline{3-10}
&&\multicolumn{4}{c}{Site (4$e$)}&&&\multicolumn{2}{c}{Site (2$a$)} \\
\cline{3-6}
\cline{9-10}
&&Ni(1)-1 & Ni(1)-2 & Ni(1)-3 & Ni(1)-4&&&Ni(2)-1&Ni(2)-2\\
\hline
$\Gamma_1^1$ & $\Psi_1$ &(100)&(-100)&(100)&(-100)&&&(200)&(-200)\\
 & $\Psi_2$ &(010)&(010)&(010)&(010)&&&(020)&(020)\\
 & $\Psi_3$ &(001)&(00-1)&(001)&(00-1)&&&(002)&(00-2)\\
\\
$\Gamma_2^1$ &$\Psi_1$ &(100)&(-100)&(-100)&(100)&&&&\\
&$\Psi_2$&(010)&(010)&	(0-10)&(0-10)&&&\multicolumn{2}{c}{------}\\
&$\Psi_3$&(001)&(00-1)&(00-1)&(001)&&&&\\
\\
$\Gamma_3^1$ & $\Psi_1$ & (100)&(100)&(100)&(100)&&&(200)&(200) \\
&$\Psi_2$&(010)&(0-10)&(010)&	(0-10)&&&(020)&(0-20)\\
&$\Psi_3$&(001)&(001)&	(001)&	(001)&&&(002)&(002)\\
\\
$\Gamma_4^1$ & $\Psi_1$ &(100)&(100)&(-100)&(-100)&&&& \\
& $\Psi_2$&(010)&(0-10)	&(0-10)&(010)&&&\multicolumn{2}{c}{------}\\
& $\Psi_3$&(001)&(001)&(00-1)&(00-1)&&&&\\
\end{tabular}
\end{ruledtabular}
\end{table}

As per the Landau theory for a second-order phase transition, only one representation can be involved in a critical transition. For \CNPO, there are two magnetic sites Ni(1) and Ni(2). For Ni(1), there are four possible magnetic structures $\Gamma_1$, $\Gamma_2$, $\Gamma_3$, and $\Gamma_4$. On the other hand, for Ni(2), there are two possible magnetic structures ($\Gamma_1$ and $\Gamma_3$). Since there is only one magnetic transition in \CNPO\ (at $T$ = 16~K), the magnetic structure should correspond to a single IR; either by $\Gamma_1$ or by $\Gamma_3$; common for both the Ni(1) and Ni(2) sites [Table \ref{tab:basis_vectors}].

 The refinement of the magnetic structure was tested for both the $\Gamma_1$ and $\Gamma_3$. Only the $\Gamma_1$ yields the best fit to the observed diffraction pattern at 1.5 K. The basis vectors for $\Gamma_1$ indicate that the moment components along all the $a$, $b$, and $c$ axes are refinable. The simultaneous refinement of all three moment components yields $m_a = -0.13\pm0.06~\mu_B$, $m_b = 0.18\pm0.10~\mu_B$, and $m_c = 1.92\pm0.04~\mu_B$ for the Ni(1) site; and $m_a = 0.16\pm0.10~\mu_B$, $m_b = 0.18\pm0.10~\mu_B$, and $m_c = 2.01\pm0.06~\mu_B$ for the Ni(2) site. It is apparent that the moments are aligned predominantly along the $c$ axis. The total ordered magnetic moment values $m_{Ni(1)} = 1.98\pm0.04~\mu_B$, and $m_{Ni(2)} = 1.96\pm0.06~\mu_B$ are in good agreement with the theoretical value of the spin only ordered moment of $2S = 2~\mu_B$. The fitted pattern, as shown in Fig. \ref{Fig:refined_pattern-1p5-25K}(b), was obtained with a model of four phases (nuclear and magnetic phases for both the \CNPO\ and Ni$_3$P$_2$O$_8$ compounds). For further clarification, the pure magnetic pattern at 1.5~K (after subtraction of nuclear background at 25~K) is shown in Fig. \ref{Fig:refined_pattern-1p5-25K}(c) with the fitted pattern by a two phase model (the magnetic phases of the \CNPO\ and Ni$_3$P$_2$O$_8$ compounds). The $R_{mag}$ factor was found to be 2.96~$\%$ for the magnetic phase of \CNPO.

\begin{figure}
\includegraphics[width=87mm]{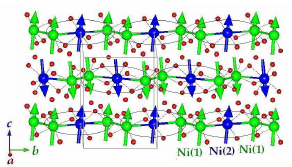}
\caption{\label{Fig:mag_struc}(Color online) The magnetic structure of \CNPO\ below $T_c$.}
 \end{figure}

The magnetic structure for \CNPO\ corresponding to the $\Gamma_1$ is shown in Fig. \ref{Fig:mag_struc}. For both of the Ni(1) and Ni(2) sites, the moment components along the $b$ axis are parallel to each other. Whereas, the moment components along the $a$ and $c$ axes have anti-parallel arrangements [Table \ref{tab:basis_vectors}]. Moreover, the resulted total moments ($m$) of the Ni(1) and Ni(2) sites arrange almost parallel to each other. This results into a ferromagnetic type trimer with an almost parallel arrangement of Ni(1)-Ni(2)-Ni(1) moments. Such trimers are coupled ferromagnetically along the chain ($b$ axis). The magnetic moment components along the $a$ and $c$ axes are coupled antiferromagnetically; implying an AFM interchain interaction. In this magnetic structure, within a unit cell the moment components along each of the $a$ and $c$ axes are canceled out individually for both the Ni(1) and Ni(2) sites. The parallel arrangement of the $b$ components ($m_b$) of individual Ni(1) and Ni(2) sites as well as among the two [Ni(1) and Ni(2)] sites results into a net magnetization ($\sim~1.08~\mu_B$ per unit cell). The net magnetization $\sim$~0.18~$\mu_B$/Ni$^{2+}$  is in good agreement with that reported from the bulk magnetization \cite{Hase.PRB.74.024430}.

The observed FM trimers [parallel arrangement of Ni(1) and Ni(2) moments within the trimers] are in agreement with the recent first-principles calculations \cite{Majumder.PRB.91.104422} where both $J_1$ and $J_2$ are estimated to be FM in nature. However, this magnetic structure is inconsistent with the predicted AFM intra-trimer interaction $J_1$ from a MC analysis of magnetization data \cite{Hase.PRB.74.024430}. The FM trimers in \CNPO\ are similar to that reported for the isostructural compound SrCo$_3$P$_4$O$_{14}$ from a neutron diffraction study \cite{Hase.JPSJ.81.064702}. A model with AFM intratrimer spin structure [antiparallel arrangement of the Ni(1) and Ni(2) spins] does not fit the observed magnetic diffraction pattern as shown by the dashed curve in Fig. \ref{Fig:refined_pattern-1p5-25K}(c). The present result of FM trimers is also consistent with the Goodenough-Kanamori rules of superexchange interactions \cite{Goodenough.PR.100.564,Kanamori.JPCS.10.87}. As per the rule, a superexchange interaction involving an exchange path with 90$^\circ$ bond angles, as in the case of \CNPO\ where the intratrimer interactions $J_1$ consist of superexchange paths Ni(1)--O2--Ni(2)/Ni(1)--O6--Ni(2) with bond angles close to 90$^\circ$ [Table \ref{T:bonds}], is preferably FM type.

\subsubsection{Critical exponent}

\begin{figure}
\includegraphics[width=70mm]{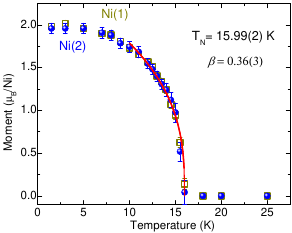}
\caption{\label{Fig:moment-T}(Color online) Temperature dependent total ordered magnetic moments for both the Ni(1) and Ni(2) sites obtained from the Rietveld analysis of the magnetic Bragg peaks. The solid line is the power law fit (Eq. \ref{eq:moment_T}) to the data. }
 \end{figure}

To investigate further the nature of the magnetic ordering, we have carried out a temperature dependent neutron diffraction study. With the increasing temperature up to the $T_c$ = 16~K, the intensity of the magnetic peaks decreases monotonically without any change in the magnetic peak positions demonstrating that there is no change of the ${\mathbf k}$-vector. The temperature dependences of the ordered moments for both the Ni(1) and Ni(2) sites are shown in Fig. \ref{Fig:moment-T}. The ordered moments were derived from the Rietveld refinement of the neutron diffraction patterns at individual temperatures. This method considers only the magnetic Bragg peaks intensities, correspond to the long-range magnetic ordering, for the determination of the magnetic moments. These measurements allow us to determine the critical exponent $\beta$ for the temperature-induced phase transition from the long-range ordered state to the paramagnetic state. At a second order magnetic phase transition, like the present case, the order parameter is a continuous function of temperature, and the critical properties can be described by the critical exponent $\beta$. Here, the order parameter is the ordered moment at the magnetic site ($m$) which is zero in the disordered or paramagnetic phase, and nonzero in the magnetically ordered phase at $T \le T_c$. The order parameter $m$ can be expressed as

\begin{equation}
\label{eq:moment_T}
m(T)=A(T_c-T)^\beta
\end{equation}

\noindent where $A$ is a proportionality constant, and $T_c$ is the long-range magnetic ordering temperature. The fit of the above equation to the observed data over a limited temperature range near $T_c$ ($0.625 \le T/T_c \le 1$) is shown by the solid red curve in Fig. \ref{Fig:moment-T}. The fitted values of $\beta$ and $T_c$ are found to be 0.36 $\pm$ 0.03, and 15.99 $\pm$ 0.02 K, respectively. The fitted value of $\beta$ agrees well with the theoretically expected value of 0.367 for the 3D Heisenberg model \cite{Blundell.book}.

\subsection{Short-range magnetic correlations}

\begin{figure}
\includegraphics[width=85mm]{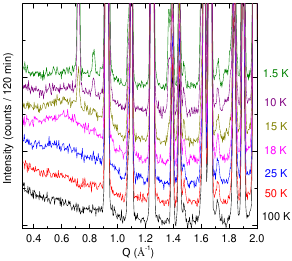}
\caption{\label{Fig:short-range}(Color online) The neutron diffraction patterns, zoomed over low $Q$ and low intensity regions, at 1.5, 10, 15, 18, 25, 50 and 100~K. The broad temperature dependent diffuse peak centered around 0.7~\AA$^{-1}$ corresponds to the short-range spin-spin correlations. The patterns are shifted vertically for better clarity.}
\end{figure}

\begin{figure} [t]
\includegraphics[width=88mm]{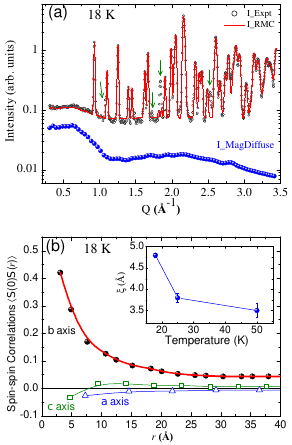}
\caption{\label{Fig:Diffuse_18K}(Color online) (a) The neutron diffraction pattern (open circles) at 18~K (above $T_N$). The solid curve through the data points is the calculated total neutron scattering by RMC method. The estimated diffuse magnetic scattering is shown by the solid spheres at the bottom. The addition peaks from the secondary phase Ni$_3$P$_2$O$_8$ are marked by arrows. (b) The spin-spin correlation functions along three crystallographic directions [$a$, $b$, and $c$ axes]. The thick solid line over the data points for $b$ axis is the fitted curve by $\sim \exp(r/\xi)$. The thin solid lines over the data points for $a$ and $c$ axes are the guide to the eyes. Inset shows the temperature dependence of the correlations length along the $b$ axis.}
\end{figure}

We now discuss the magnetic correlations above and around the $T_c$ = 16~K. The neutron diffraction patterns at 1.5, 10, 15, 18, 25, 50 and 100~K are shown in Fig. \ref{Fig:short-range}. With the lowering of temperature, a broad diffuse magnetic peak, corresponding to short-range spin-spin correlations, appears below $\sim$50~K with maximum at the $Q$ position $\sim$ 0.7~\AA$^{-1}$. This indicates that short-range spin-spin correlations start to develop at a temperature which is about three times higher than the $T_c$ =16~K. The board peak appears around the same $Q$ position of the 3D magnetic Bragg peak (0,0,1) that is found below the $T_c$ =16~K (Fig. \ref{Fig:refined_pattern-1p5-25K}). With decreasing temperature, the intensity of the broad peak increases continuously down to the $T_c$. Around the $T_c$, the broad peak gradually transforms into the magnetic sharp Bragg peak. Similar diffuse peak was reported for several quasi-1D spin-chain systems \cite{Bera.PRB.89.094402,Jain.PRB.83.184425}, including the isostructural compound \CMPO\ \cite{Hase.PRB.84.184435}, where the broad peak was assigned to one-dimension short-range spin-spin correlations. Such 1D short-range spin-spin correlations appear when the thermal energy becomes comparable to the stronger intrachain interaction in the quasi-1D compounds. At low temperature, when the weaker interchain interaction overcomes the thermal energy, the 3D long-range magnetic ordering occurs at $T_c$.

To analyze the observed weak diffuse magnetic scattering, and to investigate the nature of the short-range spin-spin correlations, we have employed the RMC method by using the program RMCPOW \cite{RMCPOW}. The main advantage of the RMC method is that it calculates total scattering which includes both total nuclear scattering (Bragg+diffuse), and total magnetic scattering (Bragg+diffuse). Hence, the individual contributions of nuclear Bragg, nuclear diffuse, magnetic Bragg and magnetic diffuse can be separated out from the total scattering in a neutron diffraction pattern. Unlike other methods, this method does not involve an analytical model for the line shape (such as, Warren \cite{Warren.PR.59.693}, Lorentzian, {\it{etc}} functions) or a {\it{priori}} assumption about magnetic order dimensionality. This method provides a useful information;  especially direct space spin-spin correlations along individual crystallographic directions. Such method has been successfully applied recently to several magnetic systems to analyze the short-range spin-spin correlations \cite{Paddison.PRL.108.017204, Nilsen.PRB.91.054415,Yu.JPCM.23.164214}. In this RMC method, the nuclear and magnetic phases were analyzed together for each of the temperature pattern where the observed intensity was corrected as per the standard data reduction method. The recorded neutron patterns were corrected for the non-sample (instrumental) background, and normalized to a vanadium standard. The overall data normalization, during the RMC refinement, was achieved by choosing a refinable scale factor for the nuclear phase, and keeping the magnetic scale factor fixed with respect to the nuclear phase. The $Q$-independent incoherent scattering from the sample was modeled by selecting a refinable flat background. The thermal diffuse (nuclear) scattering was estimated analytically from the mean square displacement parameters obtained from the Rietveld refinement (refined isotropic temperature factors, $B$). No absorption correction was considered as the overall absorption cross-section was negligible as compared to the total scattering cross-section.

For the RMC calculation at each temperature, we used the predetermined values of the lattice parameters and fractional coordinates of all atoms which were obtained from the Rietveld refinement of the diffraction patterns at the corresponding temperatures. We fixed the positions of all the atoms (i.e., without any static nuclear diffuse scattering) to accurately determine the magnetic diffuse scattering. The positions of spins were fixed at their crystallographic sites, while their orientations were refined in order to fit the data at each temperature. We also assumed that both Ni sites (2$a$ and 4$e$ sites) are equivalent magnetically, and having equal magnitude of magnetic moments. We performed the RMC simulations using a 8~$\times$~8~$\times$~8 supercell of the monoclinic unit cell containing 3072 spins. Random spin configurations with each Ni$^{2+}$ ion having effective moment $g\times \sqrt{(S(S+1))} = 2.83~\mu_B$ were taken as the starting point; and individual spin was allowed to rotate independently during the refinement. Several independent calculations were performed for each of the temperatures in order to obtain an average scattering profile.

Figure \ref{Fig:Diffuse_18K} (a) shows the experimentally observed, and the RMC calculated total scattering patterns at 18~K ($T$ $>$ $T_N$) where the intensity of the magnetic diffuse scattering is maximum. The spin-spin correlations [$\langle\roarrow{S}(0).\roarrow{S}(r)\rangle$], obtained from the RMC analysis, at 18~K are plotted along the three crystallographic directions ($a$, $b$, and $c$ axes) in real space [Fig. \ref{Fig:Diffuse_18K}(b)]. The spin-spin correlations along the chain direction ($b$ axis) are always positive (FM), and stronger. Such spatial correlations follow an exponential decay [$\sim\exp(r/\xi)$] behavior with distance $r$, as expected for ferromagnetic correlations \cite{Paddison.Science.350.179}. The spin-spin correlations along the $a$ and $c$ axes are found to be very weak. Here we would like to comment that the reported first-principles calculations \cite{Majumder.PRB.91.104422} consider the exchange interactions in the local coordination among the Ni(2) ions only [via Ni(2)-O-P1-O-Ni(2) pathway where  Ni(2)-Ni(2) distance = 6.04 ~\AA].  The calculations estimate that the $J3$ interaction (along the almost [011] direction within the $bc$ plane) is the strongest. On the other hand, in an experimental scenario (with regard to magnetic ordering), an average strength of all exchange bonds along a particular direction contributes. In the present system, the density of the $J3$ bonds is one third of that of the intrachain interactions ($J1$ and $J2$). The reduced bond density causes an effectively weaker $J3$ as compared to $J1$ and $J2$. This may result into the dominating correlation along the chain axis. For both $a$ and $c$ directions, the negative values of $\langle\roarrow{S}(0).\roarrow{S}(r)\rangle$ for the nearest neighbor distance reveal AFM correlations. The correlations are present only up to nearest neighbors. Above $T_c$, the presence of dominating spin-spin correlations along only the chain axis reveals the 1D nature of the magnetic ordering. This suggests that the effective strongest interactions are along the chain ($b$ axis) direction. The obtained spin structure at 18~K (above $T_c$), i.e., FM type moment arrangement along the chain axis, and AFM type arrangement along the $a$ and $c$ axes, is similar to that found below $T_c$=16~K for the long-range magnetic structure involving FM spin-trimers [Fig. \ref{Fig:mag_struc}]. With decreasing temperature, the short-range FM spin-spin correlation length along the chain axis increases monotonously with a correlation length of $\xi$ $\approx$ 4.8~\AA\ at 18~K [inset of Fig. \ref{Fig:Diffuse_18K}]. With decreasing temperature, a correlation develops along the chain axis when the strength of the intrachain interactions becomes comparable with the thermal energy at $\sim$~50 K. At $T_c$, the weaker interchain interactions along the $a$ and $c$ axes become comparable to the thermal energy, and a 3D long-range magnetic ordering develops.

Now we discuss below the consequence of the FM trimers in the present compound \CNPO\ on the observed magnetic properties as compared to that of the isostructural compounds in the series $AM_3$P$_4$O$_{14}$. First we focus on the 1/3 magnetization plateau that appears only for the Mn analogues, not for the Ni and Co analogues. It is reported that the 1/3 magnetization plateau for the Mn compound originates due to quantum mechanical discrete energy levels of the magnetic eigenstates of the spin-5/2 AFM trimer \cite{Hase.PRB.80.054402,Hase.PRB.84.214402}. The magnetization plateau in the magnetic field range 2--10~T appears due to an energy gap between the plateau state (total spin 5/2), and the higher state (total spin 7/2) where the increase of the magnetization with magnetic field is prevented by the energy gap. The discrete energy levels remain in SrMn$_3$P$_4$O$_{14}$ in spite of the introduction of weak 3D intertrimer interactions along the chain and between the chains. The required conditions for the 1/3 magnetization plateau were inferred as good one-dimensional characteristics and both AFM intratrimer ($J_1$) and intertrimer ($J_2$) interactions within the chains \cite{Yang.IC.47.2562}. However, the Ni and Co based-compounds have FM trimers that are coupled again ferromagnetically (both the FM $J_1$ and $J_2$) along the chain. Therefore, the present compound \CNPO\ becomes an effective $S$ =3 ferromagnetic chain, whose excitations are gapless spin waves. This is in contrast to the prediction of AFM trimer system \cite{Schmidt.JPAMT.43.405205} where the excitations are gapped. On the other hand, two magnetization plateaus at 1/3 and 2/3 of the saturation magnetization are classically expected for $S$ =1 FM trimer chains when the FM trimers ($J_1$ FM) are coupled antiferromagnetically ($J_2$ AFM), and the ratio $J_2$/$J_1$ is above a critical value \cite{Gu.JPCM.17.6081}. However, for the present compound \CNPO\, both the intratrimer interaction $J_1$, and intertrimer interaction $J_2$ are FM; hence, no magnetization plateau is observed.

In addition to the FM intratrimer and intertrimer interactions along the chain axis, the interchain interaction is also reasonably strong in \CNPO, as estimated by the first-principles calculations \cite{Majumder.PRB.91.104422}. The consequence of the strong interchain interaction in the present \CNPO\ compound is revealed by the observed higher 3D long-range magnetic ordering temperature of $T_c$=16~K as compared to the Mn-based compound ($T_N$ = 2.2~K) \cite{Hase.PRB.84.184435}. Another consequence of the stronger interchain interactions is the value of ordered moments. For the present Ni-based compound with stronger interchain interaction $J_3$; [$J_3$/(($J_1$+$J_2$)/2) $\approx$ 1] \cite{Majumder.PRB.91.104422}, the ordered moment values (1.98 and 1.96~$\mu_B$) are found to be quite close to the theoretical value of 2~$\mu_B$/Ni$^{2+}$ ($S$ =1). In spite of the spin-1 chain system, the present Ni-based compound is found to be a good realization of 3D magnetic system below $T_c$=16~K, with a negligible effect of quantum fluctuations. On the other hand, the strong effect of quantum fluctuations is present in the Mn-based ($S$ = 5/2) compound, due to weak interchain interaction (highly one dimensional magnetic lattice) [$J_3$/($J_1$+$J_2$)/2 $\ll$ 1] \cite{Hase.PRB.84.184435, Hase.PRB.80.054402}, and AFM intratrimer and intertrimer interactions along the chain; which leads to strongly reduced ordered moments of $\sim$ 3.5~$\mu_B$ (for both the sites) as compared to the theoretical value of 5~$\mu_B$/Mn$^{2+}$ ($S$ = 5/2) \cite{Hase.PRB.84.184435}. For the Co-based compound ($S$ =3/2), an intermediate effect is found where the reported experimental ordered moment values (2.9 and 2.45~$\mu_B$/Co$^{2+}$ for two sites, respectively) are slightly reduced from the theoretical value of 3~$\mu_B$/Co$^{2+}$ ($S$ =3/2) \cite{Hase.JPSJ.81.064702}.  \\

\section{Summary and conclusions}
 The microscopic spin structure, and the origin of net
magnetization in the trimer chain compound \CNPO\ have been investigated by powder neutron diffraction. With lowering of temperature, short-range spin-spin correlations develop below $\sim$50~K. With further lowering of temperature, a 3D long-range magnetic ordering sets in below $T_c$ = 16~K. Above $T_c$, the short-range magnetic ordering has 1D magnetic characteristics with dominating FM spin-spin correlations along the chain axis ($b$ axis). The nature of the magnetic structure in  both the short-range (above $T_c$) and long-range (below $T_c$) ordered states is similar. The magnetic structure is found to be consisting of FM trimers which are coupled ferromagnetically along the chain direction ($b$ axis). The moment components along the $a$ and $c$ axes are arranged antiferromagnetically. The observed magnetic structure is consistent with the first-principles calculations which estimate both the intratrimer and intertrimer interactions, $J_1$ and $J_2$, as FM. In the studied compound, a net magnetization per unit cell arises due to the uncompensated moment components along the $b$ axis, in contrast to the ferrimagnetic ordering reported earlier from the MC analysis of bulk magnetization. The present understanding of the magnetic ground state also explains the reason behind the absence of the 1/3 magnetization plateau in \CNPO\ in contrast to the isostructural Mn-compounds. The ordered moment values for both the magnetic sites [4$e$ Ni(1) and 2$a$ Ni(2)] are found to be $\sim$ 2~$\mu_B$/Ni$^{2+}$ at 1.5~K. The present system is found to be a good realization of a 3D magnetic system with $S$ = 1 trimer chains. \\

\begin{acknowledgments}
AKB acknowledges the support provided by Helmholtz-Zentrum Berlin f\"ur Materialien und Energie to carry out the neutron diffraction experiment at PSI, Switzerland.
\end{acknowledgments}

%

\end{document}